\documentclass[9pt,twocolumn,twoside]{opticajnl}
\journal{opticajournal} 

\setboolean{shortarticle}{true}

\title{Robust Characterization of Integrated Photonics Directional Couplers}

\author[1,*]{Jonatan Piasetzky}
\author[2]{Yehonatan Drori}
\author[1,2]{Yuval Warshavsky}
\author[2]{Amit Rotem}
\author[1]{Khen Cohen}
\author[1,2]{Yaron Oz}
\author[1,2]{Haim Suchowski}

\affil[1]{Raymond and Beverly Sackler School of Physics and Astronomy, Tel Aviv University, Tel Aviv, 6997801, Israel}
\affil[2]{QuanutmPulse}

\affil[*]{piasetzky1@mail.tau.ac.il}

\begin{abstract} 
Directional couplers are essential components in integrated photonics. Given their widespread use, accurate characterization of directional couplers is crucial for ensuring optimal performance. However, it is challenging due to the coupling between fibers and waveguides, which is highly sensitive to alignment and fabrication imperfections. To address these challenges, we propose a novel direct measurement technique that offers greater robustness to variations in optical interfaces, while bypassing extinction ratio measurements. Our method enables a broadband and precise characterization of the directional couplers' splitting ratio. We experimentally validate this approach, demonstrate its robustness against intentional errors, and compare it to a naive direct measurement method. Furthermore, our technique is generalized to measure the amplitude of any general 2x2 unitary circuit, providing valuable insights for designing and testing a wide range of photonic integrated devices. 

\end{abstract}

\setboolean{displaycopyright}{false} 

\begin{document}
\maketitle
\section{Introduction} 

Photonic-integrated circuits (PICs) represent a cutting-edge technology with great potential across various fields. They are pivotal for advancing on-chip optical interconnects \cite{sun_single-chip_2015}, revolutionizing data center communications \cite{vlasov_silicon_2012}, and enhancing sensing and range-finding applications \cite{xie_heterogeneous_2019}. Additionally, PICs are increasingly recognized as a promising platform for quantum technologies, including quantum information processing, quantum cryptography, and quantum sensing \cite{Wang2020, rudolph_why_2017, qiang_large-scale_2018}. The ability of PICs to integrate complex photonic functions onto a single chip makes them indispensable for these emerging technologies, underscoring the need for precise and reliable components within these systems.
\par
Among the core components of integrated photonics, directional couplers play a crucial role in manipulating light \cite{Bogaerts2012}. These devices are fundamental in splitting and combining optical signals and are integral to various applications such as wavelength division multiplexing \cite{horst_cascaded_2013}, optical switching \cite{fang_ultralow_2011}, filtering, nonlinear optics \cite{dong_observation_2006}, and quantum information processing \cite{politi_silica--silicon_2008,silverstone_-chip_2014,matthews_manipulation_2009}.

Given their widespread use and importance, the accurate characterization of directional couplers is essential for ensuring the optimal performance of photonic circuits. Measuring the performance of optical devices is crucial but challenging due to the complex coupling between the input and output fibers with the waveguides, typically achieved through grating couplers or edge coupling. These coupling coefficients significantly impact the power involved in the interaction and are highly sensitive to alignment and fabrication imperfections. Consequently, the variability in coupling location and efficiency can greatly influence measurement accuracy, making it difficult to isolate and quantify the true performance of the optical device itself. The difficulty in characterizing directional couplers leads to substantial discrepancies in performance measurement. Conventional methods for assessing these devices include direct \cite{chrostowski_silicon_2015} and indirect approaches like asymmetric Mach-Zehnder interferometers (AMZI) \cite{lu_broadband_2015, tran_robust_2016} and ring resonators \cite{katzman_robust_2022}). Direct methods often assume identical insertion losses across all ports, which is not always feasible. Indirect methods, while useful, introduce additional complexities such as enlarged circuit size, which causes additional noise and rely on noisy extinction ratio measurements. Also, these measurement characterizations can be sensitive to the wavelength and are not always broadband.
\par
To address these challenges, in this letter, we propose a novel direct measurement technique that eliminates the need for extra test structures, bypasses extinction ratio measurement, and offers greater robustness to variations in optical interfaces. This approach aims to provide a more accurate and practical means of evaluating the performance of directional couplers in integrated photonic circuits. We show the robustness of the method by applying common alignment errors. Moreover, we compare it to the more naive direct method that currently exists to show its advantage. Last, we demonstrate the methods used in characterizing couplers of different shapes, both phase matched and phase mismatched. 

\section{Theoretical backgroud}
\subsection{Directional Couplers}
\begin{figure}[H]
\centering
\includegraphics[width=\linewidth]{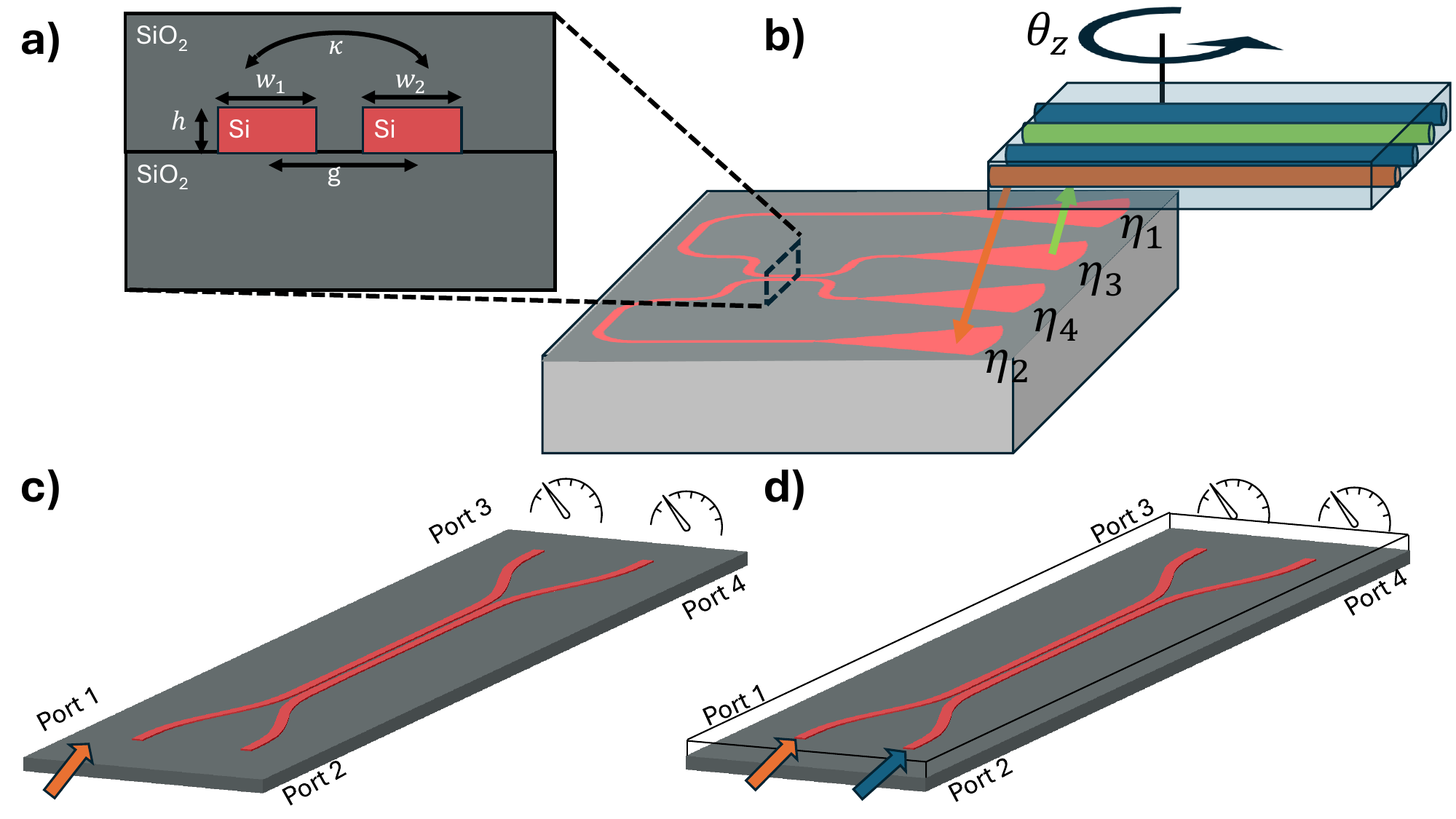}
\caption{Directional Couplers Measurements. \textbf{(a)} Cross-section of a directional coupler, composed of two evanesecencely coupled silicon waveguides. \textbf{(b)} Test structures produced. Directional couplers are coupled to a fiber array, for optical input and output, through grating couplers. In this example, light is injected through port 2 (orange) and measured from port 4 (green). \textbf{(c)} The naive direct measurement method, where light is injected through a single input port and collected at both outputs. \textbf{(d)} Our method, in which light is injected through both input ports, successively, and collected at both outputs. Thus, utilizing the unitary propagation redundancy to cancel out errors originating from optical input and output. Both the naive and our method may use a broadband source and an optical spectrum analyzer, allowing the measurement of a broad spectrum in a single shot.}
\label{fig:cells}
\end{figure}

Directional couplers consist of two closely spaced waveguides that interact through evanescent field coupling. 
Fig. \ref{fig:cells} shows such a device, with the coupling coefficient marked as $\kappa$. The geometric properties that affect the coupling coefficient are shown in the figure as well, namely the waveguide width $w_1$, $w_2$, the silicon height, $h$, and the gap between the waveguides $g$. The physics of these devices is elegantly described by coupled-mode theory. As light propagates through the coupler, energy can be transferred between the two waveguides due to the overlap of their evanescent fields. This interaction is characterized by a coupling coefficient and mode-mismatch parameters, which depend on the waveguide geometry and separation. The behavior of a directional coupler can be represented by a unitary propagator that relates the output complex-valued mode amplitudes to the input mode amplitudes. This propagator, up to a constant phase, takes the form of a 2x2 unitary matrix with elements determined by the generalized coupling strength and propagation length \cite{yariv_photonics_2007, chrostowski_silicon_2015, lifante_integrated_2003}. The mathematical description that captures the essence of the coupler's operation is given by:
\begin{multline}
\label{eq:unitary}
\begin{pmatrix}
A_3(z=L) \\ A_4(z=L)
\end{pmatrix} =  U
\begin{pmatrix}
A_1(z=0) \\ A_2(z=0)
\end{pmatrix}
\\
U \triangleq
\begin{pmatrix}
\cos(\gamma L) - i\frac{\Delta\beta}{\gamma}\sin(\gamma L) & - i\frac{\kappa}{\gamma}\sin(\gamma L) \\ 
- i\frac{\kappa}{\gamma}\sin(\gamma L) & \cos(\gamma L) + i\frac{\Delta\beta}{\gamma}\sin(\gamma L)
\end{pmatrix}
\end{multline}
Where $A_i$ are the mode amplitudes at port $i$, $\kappa$ is the coupling coefficient,  $\Delta\beta$ is the phase mismatch,$L$ is the length of the coupling region, and $\gamma \triangleq \sqrt{\kappa^2 + \Delta\beta^2}$ is the generalized coupling coefficient. In our analysis, we assume the waveguides support only a single mode, and we will normalize the mode amplitudes by the total intensities, such that $|A_1|^2+|A_2|^2=|A_3|^2+|A_4|^2=1$.
The power splitting ratio, t, is given by:
\begin{equation}
\label{eq:kappa_rabi}
    t = |U_{14}|^2 = |U_{23}|^2 = \frac{\kappa^2}{\kappa^2+\Delta\beta^2} \; sin^2\left(\gamma L\right)
\end{equation}

The importance of this formalism lies in its ability to capture the essence of the coupler's operation, including power transfer ratios and phase relationships between the output ports, making it an invaluable tool for designing and analyzing integrated photonic circuits. It allows to describe a wide range of coupling scenarios through simple variations in its parameters. For instance, adjusting the coupling coefficient allows for the modeling of different coupling strengths, while changes in the propagation length can simulate various coupler geometries. Interestingly, the dynamics described by this propagator are completely analogous to Rabi oscillations in atomic systems coupled to a near-resonant classical external field. This analogy extends further: if a phase mismatch exists between the waveguides, it manifests as a reduction in the amplitude of power oscillations, mirroring the behavior of detuned atomic systems. 

\section{Robust Measurement of the Splitting ratio}
Let us define $P_{io}$ as the measured power of light incident on port i at the output o. The measured output value depends not only on the input power, and the power splitting ratio of the directional coupler, but also on the coupling efficiencies between the grating couplers and the input and output fibers. In the most 
general form, the power measured for each input and output  is given by:
\begin{equation}\label{eq:sr}
    P_{io} = \eta_{i}\eta_{o} |U_{io}|^2 P_i
\end{equation}
Where $P_{io}$ is the measured power of light incident on port $i$ at the output port $o$, $P_i$ is the input power, $U_{io}$ is the matrix element of the directional coupler from port $i$ to port $o$, and $\eta_i$ is the coupling efficiency between grating coupler $i$ and fiber $i$ in the array. $\eta_i$ also incorporates all the constant losses that the light accumulates during its propagation in the waveguides between the directional coupler and the grating couplers, as well as the measurement efficiency (the ratio between the number of photons that populate the fiber mode and the number of photoelectron events counted).  One of the major obstacles in estimating $U_{io}$, which is the quantity we would like to measure, is the variance in $\eta_i$ between the different ports, due to imperfect alignment and variance in the insertion loss due to imperfect fabrication.  This is the reason that the naive estimation of the splitting ratio, given by $P_{14}/(P_{13} + P_{14})$, cannot be used for proper estimation of the splitting ratio.
\par
To overcome this obstacle, we present the following measurement procedure that holds when assuming negligible loss in the coupling region itself, which translates to unitary propagation of light in the directional coupler, as shown in Eq. (\ref{eq:unitary}):
\begin{equation}
\begin{aligned}
|U_{14}|^2 &= |U_{23}|^2 = t \\
|U_{13}|^2 &= |U_{24}|^2 = (1-t) \\ 
\end{aligned}
\end{equation}
One can easily see, from Eq. (\ref{eq:sr}), that the following ratio eliminates the dependence on $\eta_i$ at all, and gives a very robust estimation of the power splitting ratio of the coupler:
\begin{equation}\label{eq:kappa}
    \sqrt{\frac{P_{13}P_{24}}{P_{14}P_{23}}} = \frac{t}{1-t}
\end{equation}

Thus, to best mitigate the effect of variance in the coupling efficiency between the different fibers and grating couplers in the array, one should measure four spectral responses (from both inputs to both outputs), and extract the power splitting ratio. 

\section{Fabrication and Results}
To validate our novel measurement method, we have designed a Silicon photonic chip that contains multiple directional couplers. The chip was fabricated by Applied Nanotools, (ANT, Canada), in a multiproject Wafer run of Silicon on Insulator (SOI) with a Silicon height of 220 nm. The directional couplers were connected to a grating couplers array on the chip which was coupled to a fiber array of 16 fibers with 127 $\mu m$ pitch. 
\begin{figure}[H]
\centering
\includegraphics[width=\linewidth]{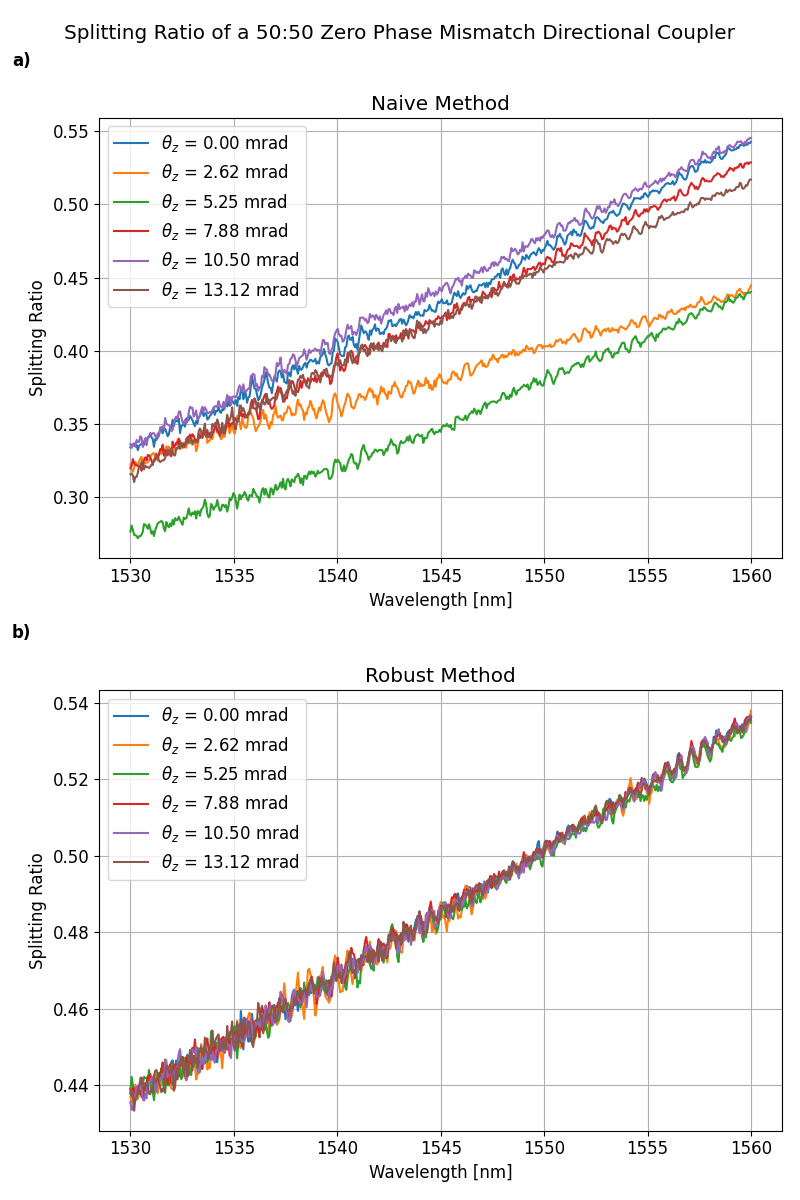}
\caption{Splitting ratio spectrum of a 50:50 phase-matched directional coupler ($X^{1/2}$ Single qubit gate in the dual rail representation) for various Yaw ($\theta_z$) angle rotations. \textbf{(a)} shows the splitting ratio, as extracted using the naive method. A considerable variance between measurements is present. \textbf{(b)} shows the splitting ratio extracted by the robust method, exhibiting no observable difference between the measurements.}
\label{fig:res}
\end{figure}

This setup is schematically shown in Fig. \ref{fig:cells}. To measure each spectrum, $P_{io}(\omega)$, an Amplified Spontaneous Emission from an Erbium-doped fiber amplifier (EDFA) source was injected into port $i$. The output at port $o$ was measured using a Yokogawa AQ6370D Optical Spectrum Analyser.  An electrically controlled fiber-coupled optical switch was used to measure all four permutations of input and output without changing the coupling insertion loss. The coupling ratio spectrum was calculated using Eq. (\ref{eq:kappa}) from these four measurements.
To verify the method's robustness, we measured the same directional coupler multiple times and compared the results to a naive estimation of the splitting ratio. We also intentionally added errors in alignment by introducing yaw rotation, which is rotation of the fiber array around the z-axis, as shown in Fig. \ref{fig:cells}\textbf{b}. This type of alignment angle error can easily arise while manually aligning the experimental setup and causes degradation of the naive measurement due to the non-uniform change in $\eta_{io}$'s. The results are shown in Fig \ref{fig:res}, it can be easily seen the robustness of the robust method (Fig. \ref{fig:res}\textbf{a}) as opposed to the naive method (Fig. \ref{fig:res}\textbf{b}). 
\begin{figure}[H]
    \centering
    \includegraphics[width=\linewidth]{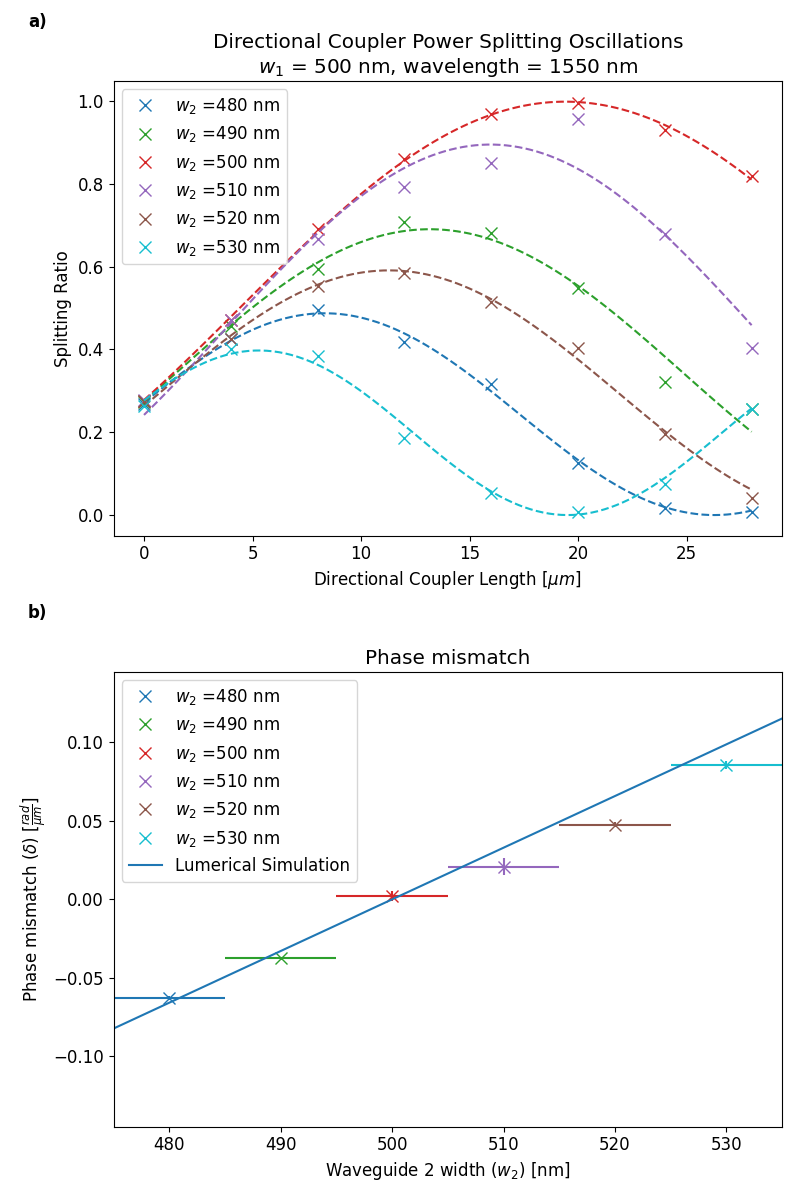}
    \caption{\textbf{(a)} Directional coupler power splitting oscillations. The width difference between the coupler's waveguides creates phase mismatch which is manifested in the oscillations as reduced amplitude, and shorter inversion length. \textbf{(b)} The extracted phase mismatch and its comparison to phase mismatch simulation done in Lumerical software.}
    \label{fig:rabi}
\end{figure}
To further evaluate our method, we have measured the spectrum of various directional couplers. All of them had the width of one waveguide fixed at 500 nm, while the width of the second waveguide varied from 480 nm to 530 nm. For each width, we have swept the length of the coupling region. The results were fitted to the power splitting ratio of non-zero phase mismatch directional couplers, as predicted by Eq. (\ref{eq:kappa_rabi}). The measured results at a wavelength of 1550 nm, along with the fitted power splitting oscillations are presented in Fig. \ref{fig:rabi}. It is clear to see that, as expected, in couplers with phase mismatch ($w_2 \ne 500 nm$), the amplitude of the splitting ratio oscillations is smaller than one, and the generalized coupling coefficient increases, as can be seen by the smaller inversion length. The couplers' phase mismatch was then extracted from the fitted parameters, and then compared to simulations performed using the commercially available photonics simulation software Lumerical. Error bars were added according to the fabrication process specification, which was given to be $\pm 5 nm$ on feature sizes. The good fit between simulation and experimental results indicates that the method proposed here performs well even for directional couplers with significant phase mismatch. 
\par
\begin{figure}[h]
    \centering
    \includegraphics[width=\linewidth]{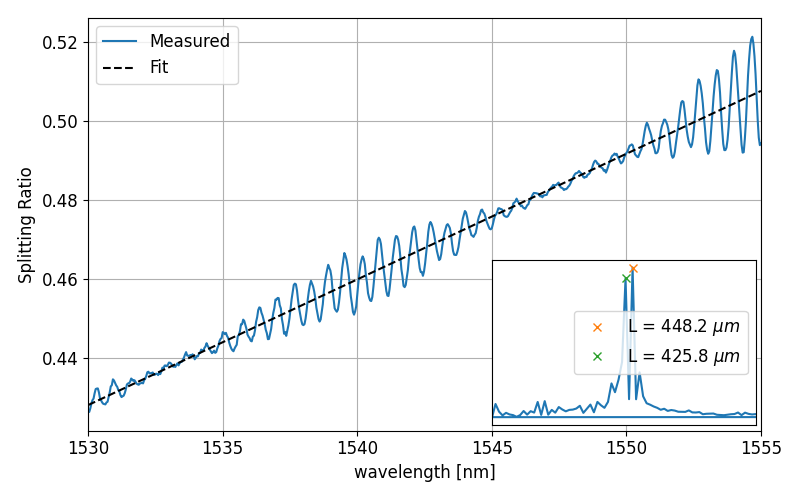}
    \caption{The spectrum of a 50:50 directional coupler. Ripples and beating are clearly present on top of a linear fit. The inset displays the signal's Fourier analysis, showing the two beating frequencies that match Fabry-Perot cavity lengths of 448.2 $\mu m$ and 425.8 $\mu m$.}
    \label{fig:fourier}
\end{figure}

The resulting coupling ratio spectrums exhibit ripples, probably attributed to standing waves caused by reflections from the coupling between the waveguide modes and the grating couplers modes, an effect that was shown in Ref. \cite{chrostowski_silicon_2015}. Furthermore, we can notice a beating effect of the ripples, which we assumed was caused by the two different, but very close, resonators (i.e., Port 1 to Port 3 and Port 1 to Port 4). To verify this assumption we performed Fourier analysis and found the two frequencies modulating the signal. They match a Fabry-Perot resonator length of 448.2 $\mu m$ and  425.8 $\mu m$, which agrees with the waveguide lengths between the grating couplers and the directional coupler, explaining the measured effect. This analysis is shown in Fig. \ref{fig:fourier}.

\section{Conclusion}
In this study, we introduce a method for measuring the splitting ratio of symmetrical and asymmetrical directional couplers, effectively mitigating the impact of alignment and fabrication errors. These errors, which often arise from the coupling between fibers and waveguides, can significantly affect characterization accuracy. Our novel direct measurement technique eliminates the need for extra test structures and bypasses extinction ratio measurement, offering greater robustness compared to traditional methods. We validated our approach using various silicon photonics-based directional couplers in both symmetrical and phase-mismatched configurations, demonstrating its resilience against intentional alignment errors. We also identified that common measurement errors from test structures can be mitigated with narrowband filtering. Although our method was designed for the characterization of directional couplers, due to the fact it relies only on the assumption that the device can be described as a unitary matrix, it can also apply to any reversible photonic device without losses. Future research could extend this method to generic 2x2 photonic circuits and even higher dimensional waveguide couplers. Our mathematical framework not only provides a practical tool for device design but also offers insights into the fundamental physics of coupled waveguide systems, bridging integrated photonics and quantum optics. Thus, we believe this method will significantly aid in the design and characterization of photonic devices for both classical and quantum applications. 

\begin{backmatter}
\bmsection{Funding} Israel Sceince Foundation (2312/21).
\bmsection{Acknowledgments} We would like to thank QuantTAU, Tel Aviv University Center for quantum science and technology, for their support.
\bmsection{Disclosures} HW: QuantumPulse (E), YO: QuantumPulse (E), YD: QuantumPulse (E), AR: QuantumPulse (E), YW: QuantumPulse (E).
\bmsection{Data availability} Data underlying the results presented in this paper are not publicly available at this time but may be obtained from the authors upon reasonable request.
\end{backmatter}

\bibliographyfullrefs{references}


\begin{thebibliography}{10}
\newcommand{\enquote}[1]{``#1''}

\bibitem{sun_single-chip_2015}
C.~Sun, M.~T. Wade, Y.~Lee, \emph{et~al.}, \enquote{Single-chip microprocessor that communicates directly using light,} {\protect\JournalTitle{Nature}} \textbf{528}, 534--538 (2015). Publisher: Nature Publishing Group.

\bibitem{vlasov_silicon_2012}
Y.~A. Vlasov, \enquote{Silicon {CMOS}-integrated nano-photonics for computer and data communications beyond {100G},} {\protect\JournalTitle{IEEE Communications Magazine}} \textbf{50} (2012).

\bibitem{xie_heterogeneous_2019}
W.~Xie, T.~Komljenovic, J.~Huang, \emph{et~al.}, \enquote{Heterogeneous silicon photonics sensing for autonomous cars [{Invited}],} {\protect\JournalTitle{Optics Express}} \textbf{27}, 3642--3663 (2019). Publisher: Optica Publishing Group.

\bibitem{Wang2020}
J.~Wang, F.~Sciarrino, A.~Laing, and M.~G. Thompson, \enquote{Integrated photonic quantum technologies,} {\protect\JournalTitle{Nature Photonics}} \textbf{14}, 273--284 (2020).

\bibitem{rudolph_why_2017}
T.~Rudolph, \enquote{Why i am optimistic about the silicon-photonic route to quantum computing,} {\protect\JournalTitle{APL Photonics}} \textbf{2} (2017). ArXiv: 1607.08535.

\bibitem{qiang_large-scale_2018}
X.~Qiang, X.~Zhou, J.~Wang, \emph{et~al.}, \enquote{Large-scale silicon quantum photonics implementing arbitrary two-qubit processing,} {\protect\JournalTitle{Nature Photonics}} \textbf{12}, 534--539 (2018). ArXiv: 1809.09791 Publisher: Nature Publishing Group.

\bibitem{Bogaerts2012}
W.~Bogaerts, P.~de~Heyn, T.~van Vaerenbergh, \emph{et~al.}, \enquote{Silicon microring resonators,} {\protect\JournalTitle{Laser and Photonics Reviews}} \textbf{6}, 47--73 (2012).

\bibitem{horst_cascaded_2013}
F.~Horst, W.~M.~J. Green, S.~Assefa, \emph{et~al.}, \enquote{Cascaded {Mach}-{Zehnder} wavelength filters in silicon photonics for low loss and flat pass-band {WDM} (de-)multiplexing,} {\protect\JournalTitle{Optics Express}} \textbf{21}, 11652--11658 (2013). Publisher: Optica Publishing Group.

\bibitem{fang_ultralow_2011}
Q.~Fang, J.~F. Song, T.-Y. Liow, \emph{et~al.}, \enquote{Ultralow {Power} {Silicon} {Photonics} {Thermo}-{Optic} {Switch} {With} {Suspended} {Phase} {Arms},} {\protect\JournalTitle{IEEE Photonics Technology Letters}} \textbf{23}, 525--527 (2011). Conference Name: IEEE Photonics Technology Letters.

\bibitem{dong_observation_2006}
P.~Dong, J.~Upham, A.~Jugessur, and A.~G. Kirk, \enquote{Observation of continuous-wave second-harmonic generation in semiconductor waveguide directional couplers,} {\protect\JournalTitle{Optics Express, Vol. 14, Issue 6, pp. 2256-2262}} \textbf{14}, 2256--2262 (2006). Publisher: Optica Publishing Group.

\bibitem{politi_silica--silicon_2008}
A.~Politi, M.~J. Cryan, J.~G. Rarity, \emph{et~al.}, \enquote{Silica-on-silicon waveguide quantum circuits,} {\protect\JournalTitle{Science}} \textbf{320}, 646--649 (2008). ArXiv: 0802.0136 Publisher: American Association for the Advancement of Science.

\bibitem{silverstone_-chip_2014}
J.~W. Silverstone, D.~Bonneau, K.~Ohira, \emph{et~al.}, \enquote{On-chip quantum interference between silicon photon-pair sources,} {\protect\JournalTitle{Nature Photonics}} \textbf{8}, 104--108 (2014). ArXiv: 1304.1490 Publisher: Nature Publishing Group.

\bibitem{matthews_manipulation_2009}
J.~C. Matthews, A.~Politi, A.~Stefanov, and J.~L. O'Brien, \enquote{Manipulation of multiphoton entanglement in waveguide quantum circuits,} {\protect\JournalTitle{Nature Photonics}} \textbf{3}, 346--350 (2009). ArXiv: 0911.1257.

\bibitem{chrostowski_silicon_2015}
L.~Chrostowski and M.~E. Hochberg, \emph{Silicon photonics design} (Cambridge University Press, Cambridge, United Kingdom, 2015). OCLC: 906944188.

\bibitem{lu_broadband_2015}
Z.~Lu, H.~Yun, Y.~Wang, \emph{et~al.}, \enquote{Broadband silicon photonic directional coupler using asymmetric-waveguide based phase control,} {\protect\JournalTitle{Optics Express}} \textbf{23}, 3795 (2015).

\bibitem{tran_robust_2016}
M.~A. Tran, T.~Komljenovic, J.~C. Hulme, \emph{et~al.}, \enquote{A {Robust} {Method} for {Characterization} of {Optical} {Waveguides} and {Couplers},} {\protect\JournalTitle{IEEE Photonics Technology Letters}} \textbf{28}, 1517--1520 (2016). Publisher: Institute of Electrical and Electronics Engineers Inc.

\bibitem{katzman_robust_2022}
M.~Katzman, Y.~Piasetzky, E.~Rubin, \emph{et~al.}, \enquote{Robust {Directional} {Couplers} for {State} {Manipulation} in {Silicon} {Photonic}-{Integrated} {Circuits},} {\protect\JournalTitle{Journal of Lightwave Technology}}  (2022). Publisher: Institute of Electrical and Electronics Engineers Inc.

\bibitem{yariv_photonics_2007}
A.~Yariv, P.~Yeh, and A.~Yariv, \emph{Photonics: optical electronics in modern communications}, The {Oxford} series in electrical and computer engineering (Oxford University Press, New York, 2007), 6th ed. OCLC: ocm58648003.

\bibitem{lifante_integrated_2003}
G.~Lifante, \emph{Integrated {Photonics} {Fundamentals}} (Wiley, 2003).

\end{thebibliography}
\end{document}